\documentclass[sigconf]{acmart}
\AtBeginDocument{%
  \providecommand\BibTeX{{%
    \normalfont B\kern-0.5em{\scshape i\kern-0.25em b}\kern-0.8em\TeX}}}

\copyrightyear{2022}
\acmYear{2022}
\setcopyright{acmcopyright}\acmConference[SIGIR '22]{Proceedings of the 45th
International ACM SIGIR Conference on Research and Development in Information
Retrieval}{July 11--15, 2022}{Madrid, Spain}
\acmBooktitle{Proceedings of the 45th International ACM SIGIR Conference on
Research and Development in Information Retrieval (SIGIR '22), July 11--15, 2022,
Madrid, Spain}
\acmPrice{15.00}
\acmDOI{10.1145/3477495.3531869}
\acmISBN{978-1-4503-8732-3/22/07}

\usepackage{comment}
\usepackage{enumerate}
\usepackage{balance}
\usepackage[shortlabels]{enumitem}
\usepackage[singlelinecheck=false,justification=justified]{caption}
\usepackage{algorithm} 
\usepackage{algpseudocode} 
\usepackage{amsmath}

\usepackage{amssymb}
\usepackage{multirow}
\usepackage{float}
\usepackage{lipsum}
\DeclareMathOperator*{\argmax}{argmax}
\usepackage{graphics}
\usepackage{tikz,pgfplots}
\usepackage{soul}
\usepackage{capt-of}
\usepackage{caption}
\usepackage{soul}
\usepackage{makecell}
\usepackage{bm}
\usepackage{dblfloatfix}
\usepackage{natbib}
\usepgfplotslibrary{groupplots}

\newcolumntype{Y}{>{\centering\arraybackslash}X}
\usepackage{tabularx}
\usepackage{amsthm}

\usepackage{underoverlap}
\usepackage{lscape}
 \newcolumntype{b}{>{\hsize=1.6\hsize}X}

\usepackage{kotex}

\definecolor{customblue}{HTML}{20639B}
\definecolor{customred}{HTML}{ED553B}
\definecolor{customyellow}{HTML}{EB9605}
\definecolor{custompurple}{HTML}{9867C5}
\definecolor{customgreen}{HTML}{35B37E}

\definecolor{shcolor}{rgb}{0,0,1}

\begin{document}
\fancyhead{}

%\title{MOI-Mixer: Multi-Order-Interaction based Mixer \\ for Sequential Recommendation}
\title{Towards Validating Long-Term User Feedbacks \\in Interactive Recommendation Systems }

\settopmatter{authorsperrow=4}
\author{Hojoon Lee}
% \authornote{*Equal Contribution. †Corresponding author.}
%\authornote{*Equal Contribution.}
\email{joonleesky@kaist.ac.kr}
%\email{webmaster@marysville-ohio.com}
\affiliation{%
  \institution{KAIST}
  %\streetaddress{P.O. Box 1212}
  \country{}
}

\author{Dongyoon Hwang}
\email{godnpeter@kaist.ac.kr}
\affiliation{%
  \institution{KAIST}
  \country{}
} 

\author{Kyushik Min}
\email{leonard.q@kakaoenterprise.com}
\affiliation{%
  \institution{KAKAO Enterprise}
  \country{}
} 

\author{Jaegul Choo}
\email{jchoo@kaist.ac.kr}
\affiliation{
  \institution{KAIST}
  \country{}
} 

\renewcommand{\shortauthors}{Lee et al.}





%% page


%%%%%% BK variables %%%%%%

\begin{abstract}
Interactive Recommender Systems (IRSs) have attracted a lot of attention, due to their ability to model interactive processes between users and recommender systems. Numerous approaches have adopted Reinforcement Learning (RL) algorithms, as these can directly maximize users’ cumulative rewards. In IRS, researchers commonly utilize publicly available review datasets to compare and evaluate algorithms. However, user feedback provided in public datasets merely includes instant responses (e.g., a rating), with no inclusion of delayed responses (e.g., the dwell time and the lifetime value). 
Thus, the question remains whether these review datasets are an appropriate choice to evaluate the long-term effects in IRS.  
In this work, we revisited experiments on IRS with review datasets and compared RL-based models with a simple reward model that greedily recommends the item with the highest one-step reward. Following extensive analysis, we can reveal three main findings: First, a simple greedy reward model consistently outperforms RL-based models in maximizing cumulative rewards. Second, applying higher weighting to long-term rewards leads to degradation of recommendation performance. Third, user feedbacks have mere long-term effects in the benchmark datasets. Based on our findings, we conclude that a dataset has to be carefully verified and that a simple greedy baseline should be included for a proper evaluation of RL-based IRS approaches.
\end{abstract}

\maketitle

\section{Introduction} \label{section:intro}
Over the past decades, we have witnessed a huge growth in recommender systems in various e-commerce platforms and social network services ~\cite{2009Koren-MatrixFactorization, Sarwar2001-item_CF, sasrec, bert4rec, Covington2016-youtube}. 
However, conventional recommender systems regard recommendation as a one-step decision-making task and do not consider future changes in user preferences. This prohibits the recommender system from further optimizing users’ long-term satisfaction. As a response, Interactive Recommender Systems (IRSs) have been proposed to reformulate the recommendation scenario as a multi-step decision-making process. 
In contrast to a conventional recommendation scenario, the goal of an IRS is to provide recommendations that can maximize a user’s cumulative satisfaction.

%due to its ability to model the sequential interactive process between the user and the recommendation system.
%Unlike conventional recommendation systems which treat recommendations as a one-step prediction task, IRS reformulates the recommendation scenario as a multi-step decision-making process.
%Intuitively, applying Reinforcement Learning (RL) algorithms for the IRS recommendation scenario has attracted a lot of research from the research community.
%A lot of community effort has been devoted to develop an effective algorithms for IRS. 
%Since the Reinforcement Learning itself is designed to

Reinforcement learning (RL) algorithms, whose objective is to optimize long-term rewards, have naturally attracted much interest in the interactive recommendation scenarios and have been successfully applied in real-world applications~\cite{chen2019top, zhao2020jointly, zhao2018dqn}. 
Not only have they been able to maximize the immediate user feedbacks (e.g., clicks), they have also been able to maximize the delayed feedbacks (e.g., lifetime value) which is not apparent at first, but which can be profitable in the long run.
%Reinforcement Learning (RL) algorithms were the key ingredients in maximizing the cumulative user satisfaction. %such as the overall dwell-time or lifetime value. 
However, since the experiments are conducted on their private dataset, researchers commonly rely on public review datasets (e.g., MovieLens) to compare and develop the algorithms.
Unlike the private dataset, public review datasets only contains instant feedbacks (e.g., movie rating) without any inclusion of the delayed feedbacks (e.g., lifetime value) from the users.
Therefore, this raises questions around the significance of long-term effects in the review datasets.
If the long-term effect is insignificant, a simple recommender system that maximizes the one-step reward will be sufficient, making the dataset inappropriate to benchmark the IRS.

%지금 당장 one-step만을 고려하는 추천으로 충분히 해결이 가능하고 IRS formulation에 적합하지 않다.
%the optimal choice for evaluating RL-based methods which aim to maximize the long-term cumulative rewards.
%public review datasets only include an instant user feedbacks (e.g., rating), and do not include delayed feedbacks (e.g., dwell-time, life time value).
%This raises question whether review datasets are the optimal choice for evaluating RL-based methods which aim to maximize the long-term cumulative rewards.
%do actually reflect any long-term utility related user behaviours which RL-based methods can exploit.
%Efforts have been made on effectively extracting users’ long-term interests from such log data has become a key challenge for constructing an effective recommendation system.
%Currently, there is an active field of research which involves both model-based and model-free RL techniques where various successful results in real-world applications including YouTube, Alibaba, etc. ahve been shown. 

% 결국 포인트는 benchmark dataset인데, 그 얘기를 어케해야할지를 잘 모르겠넹.... 흠...
In this work, we first revisited experiments involving RL-based IRSs to verify the significance of long-term effects on public datasets. 
First, we compared RL-based models with a simple recommender system that greedily recommends the item with the highest one-step reward. We found that this greedy method outperforms RL models. 
Second, we ablated through a discount factor to compare the performance of RL models according to different weights on future rewards. Here, we observed a gradual degradation in the performance when more weights were applied to future rewards. 
Finally, by comparing the approximated optimal policy and the simple greedy policy, we were able to conclude that user feedbacks in the public review datasets have insignificant long-term effects.

From these experimental results, we believe that current public review datasets may not be an appropriate choice to benchmark the RL-based IRS. Therefore, for a proper evaluation, we suggest to carefully analyze the dataset and include a simple greedy baseline to verify the effectiveness of an RL-based IRS.

%In this work, we study whether the commonly utilized benchmark review datasets are whethean appropriate choice for evaluating a recommendation model which aims to maximize the long-term accumulative user utility. We start with revisiting the experiments of RL-based recommendation papers. 
%Surprisingly, we achieve experiment results which suggest that RL-based recommendation systems are outperformed by a simple reward model which greedily recommends the item with the highest one-step reward. 
%We further investigate this issue in a second experiment and empirically show that there is a gradual degradation in the performance of each baseline recommendation system when incorporating more weight on future rewards. 
%This raises questions whether the commonly utilized review datasets are an appropriate choice when evaluating an interactive recommendation system since they show subtle signs of long-term dependencies within the user-item interactions. 
%Finally, we analyze the benchmark datasets to investigate the existence of beneficial long-term effects in the public review dataset. If none, a greedy policy would be able to match the performance of an optimal policy for the benchmark review datasets.

%provide data analysis to show that commonly utilized benchmark review datasets lack in long-term, delayed rewards (i.e., positive feedback), and that a simple greedy approach shows near optimal performance, raising question whether they should be utilized in evaluating the effectiveness of RL-based model's capacity of learning long-term rewards.

\section{Related Work} \label{section:relatedwork}
%Reinforcement learning based recommendation methods consider the recommendation procedure as an interactive process and formulate it as a Markov Decision Process (MDP).
Interactive recommender system (IRS) is designed to model the sequential interaction between a user and the recommender system. 
Traditionally, contextual bandits~\cite{Li2010-contextualbandit, Qin2014-contextualbandit, Wu2016-contextualbandit, Wang2017-contextualbandit} are used to learn the empirical utilities from the online interaction while handling the explore/exploit dilemma of the recommendations. 
However, learning directly from the online experience is expensive and may hurt user experience. Therefore, the focus has been shifted to learning the recommender system through the user’s logged behavior data.

%(i) Contextual bandits and (ii) RL-based approaches are the two main research directions of previous work on IRS.
%Contextual bandits are proposed to learn from empirical rewards received from interactions with real users and handle the explore/exploit dilemma in online recommendations~\cite{Li2010-contextualbandit, Qin2014-contextualbandit, Wu2016-contextualbandit, Wang2017-contextualbandit}.
%However, since the contextual bandit algorithms are designed 
%, which assumes that the users' interests are stationary and mainly focuses on the online recommendation setting, we do not consider this line of research in our work. 

%On the other hand, recent approaches combining neural networks, which are highly representative function approximators, with classic RL approaches, where the core idea is to learn an optimal strategy which maximizes the expected sum of rewards over time (i.e., utilities, feedback). 
%Thus, recent work on IRS mostly prefer RL-based approaches. %due to their capability of effectively representing the dynamic users' preferences and .
%The core idea of RL is to learn an optimal strategy which maximizes the cumulative long-term rewards for the users over time.

Recent works on IRS has adopted RL algorithms to identify optimal recommendation strategies from the logged behavior data.
These approaches have shown great success in real-world e-commerce and social media / networking platforms such as \textit{Alibaba} \cite{zhao2018dqn}, \textit{TikTok} \cite{zhao2020jointly}, and  \textit{YouTube} \cite{chen2019top}.
However, such work utilizes a private dataset, making them inaccessible for the research community. As an alternative, researchers commonly rely on public review datasets to compare and develop RL-based recommendation algorithms.
The RL-based IRS can be mainly categorized into two groups, policy-based and value-based methods. 
Policy-based methods attempt to learn an optimal recommendation policy by applying policy gradient to a parameterized policy \cite{liu2018drr, chen2019top, chen2019tree_pg}. 
On the other hand, value-based methods aim to learn the Q-value for each state and perform the action with the highest Q-value \cite{zhao2018dqn, zou2020nicf}. 
Actor-critic algorithms~\cite{xin2020self, zhao2018ddpg}, which integrates policy- and value-based methods have also been studied.
There, the recommendation policy serves as an “actor” and is trained to maximize the value from the trained “critic”.

\section{Problem Formulation} \label{section:rq1}

We first illustrate how the recommender system and a user jointly build an interactive process.
As the user enters the service or platform, the recommender system constructs the user's profile based on the items they have interacted with and corresponding feedback.
By inferring the user's latent interests behind the interaction history, the system provides an item to the user and receives feedback on the recommended item. 
The feedback can be either explicitly provided such as via ratings, or inferred from an implicit reaction such as views or clicks.
After receiving the feedback, the system updates the user's profile and continues to recommend the next item. 
This interaction loop proceeds until the user leaves the platform. The goal of the recommender system is to maximize the cumulative "rewards" during the interaction, which is the numerical representation of the feedback.

Following previous work \cite{wang2019mbcal, zou2020nicf, zhao2018ddpg, zhou2020knowledge_graph, chen2019tree_pg, zhao2018dqn}, this interactive process is formulated as a Markov Decision Process (MDP). 
Formally, MDP consists of a tuple of $(\mathcal{S}, \mathcal{A}, \mathcal{R}, \mathcal{P}, \gamma)$ as follows:

\renewcommand\labelitemi{\tiny$\bullet$}
\begin{itemize}[leftmargin=*]
\item $\mathcal{S}:$ a continuous state space to describe the user's profile. A state $s_t^u \in \mathcal{S}$ is defined as the previous interaction history of the user $u$ before time-step $t$. For each time-step $t$, the interaction consists of the recommended item $i_t^u$, and its corresponding feedback $f_{t}^u$.
\begin{equation}
    s_t^u = \{(i_1^u, f_{1}^u), (i_2^u, f_{2}^u), ... , (i_{t-1}^u, f_{t-1}^u)\}
\label{eq:state}
\end{equation}
In this work, we consider the user's provided rating for the recommended item $i_t^u$ as the feedback.

\item $\mathcal{A}:$ a discrete action space with candidate items to recommend. An action $a_t^u \in \mathcal{A}$ denotes the recommended item by the recommender system at time-step $t$ for the user $u$.

\item $\mathcal{R}: \mathcal{S} \times \mathcal{A} \rightarrow \mathbb{R}$ is the reward function where $r(s_t^u,a_t^u)$ denotes the immediate reward from user $u$ at state $s_t^u$ by taking action $a_t^u$. The flexibility of designing the reward function allows the integration and optimization of user's diverse feedbacks. Following from \cite{zou2020nicf, zhou2020knowledge_graph}, we set the provided rating as the reward.

\item $\mathcal{P}: \mathcal{S} \times \mathcal{A} \times \mathcal{S} \rightarrow \mathbb{R}$ is a state transition probability. 

\item $\gamma$ is a discount factor that controls the weights for future rewards.
\end{itemize}

Without loss of generality, we aim to learn the recommendation policy $\pi : \mathcal{S} \rightarrow \mathcal{A}$ which maximizes the cumulative rewards as:
\begin{equation}
    \pi^* =\argmax_{\pi} \mathbb{E}[\sum_{u=1}^U\sum_{t=1}^T  r(s_t^u,a_t^u)]
\label{eq:state}
\end{equation}
where $T$ denotes the total number of steps of the interaction process and $U$ denotes the number of all users.

\section{Revisiting IRS Experiments} \label{section:rq2}

In this section, we revisit the experiments of the IRS papers \cite{zou2020nicf, zhou2020knowledge_graph, chen2019tree_pg, zhao2018ddpg} that utilized RL algorithms into recommender system. Although these RL-based recommendation models directly optimize the long-term satisfaction of the users, we show that a simple greedy recommendation model yields competitive or even better results.

\subsection{Experimental Setup} \label{section:task_settings}

\noindent
\textbf{Datasets} We evaluate our model using four standard recommendation review datasets in IRS: EachMovie, Movielens-1M, -20M, and Netflix, where explicit ratings are available for each interaction. 
For each dataset, interacted items are grouped by users and ordered by timestamps. Following~\cite{sasrec}, we only kept users and items that have at least five interactions. %to maintain the quality of the dataset. 
Statistical deatails are outlined in Table~\ref{table:dataset}.

\begin{table}[h]
\caption{Statistics of the processed dataset. The Avg.Int denotes the average number of interacted items per user.}
\begin{tabular}{l c c c c c}
\toprule
&\\[-3ex]               
Dataset     & \# Users     & \# Items    & \# Int             & Avg.Int       \\[0.2ex]
\hline
&\\[-2ex]
EachMovie   & 56,071       & 1,613       & 2,798,088          & 49.90         \\
ML-1M       & 6,040        & 3,416       & 999,611            & 165.50        \\
ML-20M      & 138,493      & 18,345      & 19,984,024         & 144.30        \\
Netflix     & 472,987      & 17,769      & 100,461,928        & 212.40        \\[0.2ex]
\bottomrule
\end{tabular}
\label{table:dataset}
\end{table}

\begin{table*}[t]
\caption{Performance of different models on the interactive recommender system. The results are averaged over 10 random seeds. Bold scores indicate the best model for each metric and underlined scores indicate the second best model.}
\resizebox{2.0\columnwidth}{!}{
\begin{tabularx}{\textwidth}{@{}bYYYYYYYYYYYYYYY@{}} 
\toprule
&\\[-3ex]
\multirow{2}{*}{Methods} 
& \multicolumn{3}{c}{EachMovie}   & \multicolumn{3}{c}{ML-1M}  & \multicolumn{3}{c}{ML-20M} & \multicolumn{3}{c}{Netflix}   \\[-0.2ex]
 \cmidrule(lr){2-4} \cmidrule(lr){5-7}  \cmidrule(lr){8-10} \cmidrule(lr){11-13} 
 & RW@40 & PR@40 & RC@40 & RW@40 & PR@40 & RC@40 & RW@40 & PR@40 & RC@40 & RW@40 & PR@40 & RC@40 \\[0.2ex]
\hline
&\\[-2.ex]
Random   
         & 3.167         & 0.0203           & 0.0246
         & 3.241         & 0.0900           & 0.0094 
         & 2.885         & 0.0450           & 0.0010
         & 2.949         & 0.0668           & 0.0011 \\
POP      
         & 4.021         & 0.0723           & 0.1620
         & 4.353         & 0.8521           & 0.1415 
         & 4.002         & 0.5445           & 0.0802
         & 3.680         & 0.3757           & 0.0294 \\
SASRec \cite{sasrec}  
         & 3.449         & 0.0872           & 0.0677
         & 3.879         & 0.5243           & 0.0896   
         & 3.455         & 0.2235           & 0.0124
         & 3.648         & 0.3399           & 0.0195 \\[0.2ex]
\hline
&\\[-2.2ex]
%LinUCB   & -            & -             & -
%         & -            & -             & -
%         & -            & -             & -
%         & -            & -             & - \\
%HLinUCB  & -            & -             & -
%         & -            & -             & -
%         & -            & -             & -
%         & -            & -             & - \\[0.2ex]
%\hline
%&\\[-2ex]
DQNR \cite{zhao2018dqn}
         & \underline{5.079}& \underline{0.6420}& \underline{0.6552}
         & \textbf{4.689}& \textbf{0.9950}& 0.1597   
         & \underline{4.483}& \underline{0.9475}& \underline{0.1269}
         & 4.608         & 0.9530          & 0.1226 \\
NICF \cite{zou2020nicf}     
         & 5.075         & 0.6289          & 0.6434
         & 4.677         & 0.9947          & \underline{0.1662}  
         & 4.478         & 0.9421          & 0.1262
         & \underline{4.655}& \underline{0.9548}& \underline{0.1230} \\
SQN \cite{xin2020self}
         & 4.521         & 0.3734          & 0.4468
         & 4.134         & 0.6285          & 0.0662
         & 3.973         & 0.5048          & 0.0557
         & 3.497         & 0.2374          & 0.0105 \\
SAC \cite{xin2020self}
         & 4.225         & 0.1104          & 0.2352
         & 3.961         & 0.4943          & 0.0728   
         & 3.845         & 0.4047          & 0.0463
         & 3.589         & 0.2601          & 0.0119 \\
DDPGR \cite{liu2018drr}
         & 4.063         & 0.0681            & 0.0917
         & 3.750         & 0.3696            & 0.0683
         & 3.673         & 0.2370            & 0.0310
         & 3.588         & 0.0730            & 0.01389  \\[0.2ex]
\hline
&\\[-2.2ex]
GreedyRM   
         & \textbf{5.116}   & \textbf{0.6690} & \textbf{0.6578}
         & \underline{4.680}& \underline{0.9948} & \textbf{0.1695} 
         & \textbf{4.536}   & \textbf{0.9712} & \textbf{0.1318}
         & \textbf{4.689}   & \textbf{0.9596} & \textbf{0.1244} \\
\bottomrule
\end{tabularx}}
\label{table:baseline_performance_result}
\end{table*}

%\newline
%\newline
%\textbf{Average Reward@T} Since the goal of the IRS is to maximize the cumulative sum of rewards throughout the interaction, a straightforward evaluation measure is the average reward obtained by the interaction sequence with length $T$.
%\begin{equation}
%    \text{RW@T} = {1 \over {\#\text{users} \times T}}
%    \sum^{\text{\#users}}_{u=1} \sum^{T}_{t=1} r(s_t, a_t) 
%\end{equation}

%In order to estimate calculate the average cumulative precision and recall, we count the number of positive interactions between the user and generated recommendations. 
%\newline
%\textbf{Average Cumulative Precision@T}
%\begin{equation}
%    \text{PR@T} = {1 \over {\#\text{users} \times T}}
%    \sum^{\text{\#users}}_{u=1} \sum^{T}_{t=1} p_t
%\end{equation}

%\textbf{Average Cumulative Recall@T} 
%\begin{equation}
%    \text{RC@T} = {1 \over {\#\text{users} \times T}}
%    \sum^{\text{\#users}}_{u=1} \sum^{T}_{t=1} {p_t \over {\# \text{positive items}}}
%\end{equation}

%As for $p_t$, we set $p_t =1$ if $f_{u, t}$ is positive. 

\noindent
\textbf{Models} 
We compared the following representative baselines:
\renewcommand\labelitemi{\tiny$\bullet$}
\begin{itemize}[leftmargin=*]
\item Random: A model that randomly recommends the items.
\item POP: A model that recommends the most popular item. 
\item SASRec \cite{sasrec}: A uni-directional Transformer model that is trained to predict the user's next interacted item.
\item DQNR \cite{zhao2018dqn}: A DQN~\cite{mnih2015human}-based model that estimates Q-value for each action and recommends the item with the highest Q-value.
\item NICF \cite{zou2020nicf}: A model similar to DQNR where the target Q-value is computed within the user's previous interacted items.
\item SQN \cite{xin2020self}: A model that jointly optimizes the Q-value and the probability of interaction for each item.
\item SAC \cite{xin2020self}: The actor-critic version of SQN where the interaction probability is weighted by the estimated Q-value.
\item DDPGR \cite{zhao2018ddpg}: A DDPG~\cite{lillicrap2015continuous}-based model that addresses the large discrete action space by learning a ranking vector.
\item GreedyRM: Our proposed simple baseline that estimates the one-step reward for each action and greedily recommends the item with the highest reward. (i.e., DQNR with $\gamma=0$).
\end{itemize}

\noindent
\\
\textbf{Architecture} For a fair comparison between the learning algorithms, we unified the network architecture with a uni-directional Transformer as SASRec \cite{sasrec} to encode the states .

\begin{figure}[h]
\begin{center}
\includegraphics[width=0.9\linewidth]{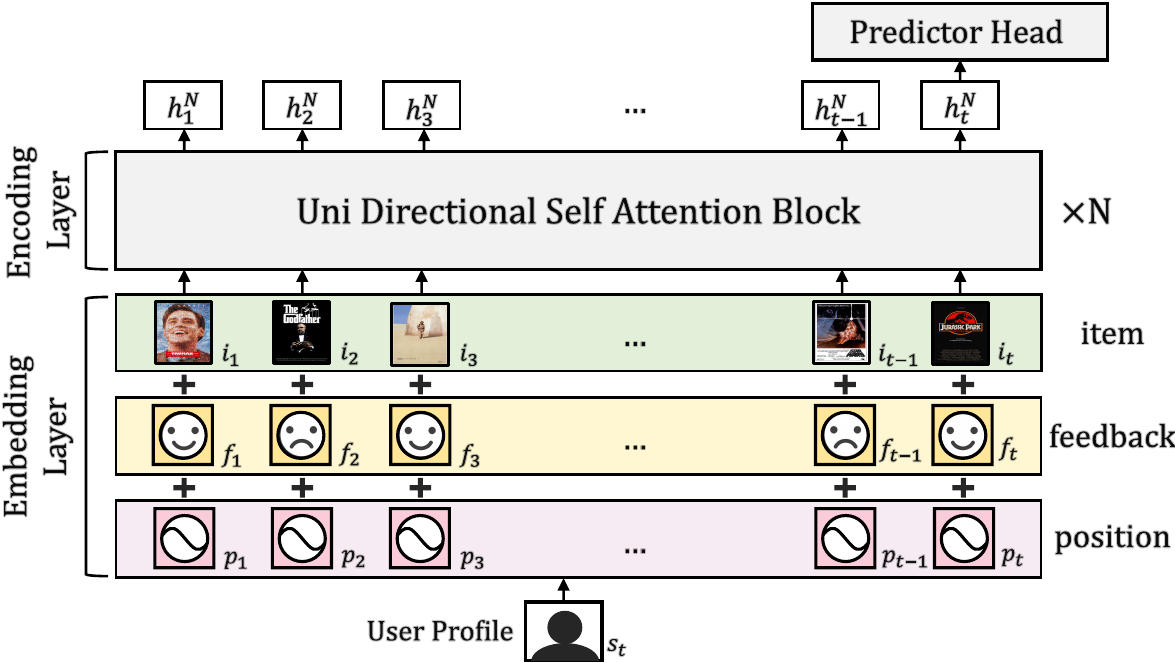}
\end{center}
\captionsetup{justification=centering}
\caption{Unified architecture for the recommendation models. Each model only differs in the Predictor Head.} 
\label{fig:architecture}
\end{figure}
\vspace{-0.3cm}
Given a state $s_t$, the input embedding is constructed by summing up the embedded items $[i_1,\text{…},i_t]$, feedbacks $[f_1,\text{…},f_t]$, and pre-defined sinusoidal positional embedding. Then, the input embedding is fed into the self-attention block to construct the hidden representation $[h_1,\text{…},h_t]$. The last hidden representation $h_t$ is passed onto the prediction head with two fully-connected layers and projected onto the item embedding space.
For each item, the projection score indicates the reward for GreedyRM, Q-value for DQNR, NICF, and un-normalized log probability in SASRec.
SQN, SAC, and DDPGR have two separate prediction heads, one for policy and another for Q-value.

\vspace{-0.1cm}
\noindent
\\
\textbf{Training} We re-implemented the aforementioned baselines by following the details of each paper. Following \cite{sasrec, zou2020nicf}, we set the number of layers as $N=2$, hidden dimension size as $d=64$, maximum sequence length as $s=200$, and set the discount factor $\gamma$ as $0.95$. For training, we used an Adam optimizer with $\beta_1=0.9$, $\beta_2=0.999$ with gradient clipping when its $l_2$ norm exceeds $5$ and set the batch size to $256$. 

For all models, we tune the hyper-parameters based on the performance by varying the learning rate $lr$ from $\{0.01, 0.001, 0.0001\}$ and $l_2$ regularization weight $wd$ from $\{0.001, 0.0001, 0.00001\}$. 

\subsection{Evaluation Protocol}

Following \cite{zou2020nicf, zhou2020knowledge_graph, chen2019tree_pg}, the datasets were split into three disjoint sets: 85\% of users for training, 5\% of users for validation, and 10\% for testing.
In the evaluation, the user profile was constructed based on their initial $40$ steps of interaction history and the interactive recommendation process was run through for $40$ time-steps to measure the recommendation performance. 

However, unless one happens to own a service or a platform, obtaining true feedback from the test user is infeasible and is also prohibitively expensive. Therefore, a common evaluation protocol is to build an environment simulator that mimics the user’s behavior and measures the performance within the learned simulator. 
Often, matrix factorization is used as the simulator to predict the user's behavior~\cite{zhou2020knowledge_graph, Hu2018-simulator, liu2018drr}. 
However, matrix-factorization is limited in modeling the dynamic changes of the user's preferences, hence produce predictions that deviate from the actual user's behaviors.
Therefore to accurately model the user's feedback, we construct the simulator based on the self-attentive architecture as Figure \ref{fig:architecture}. We report the RMSE value of our simulator for each dataset in Table \ref{table:simulator_RMSE} to assess the effectiveness of the constructed simulator,

\begin{table}[h]
\caption{RMSE value of the environment simulators.}
\resizebox{0.95\columnwidth}{!}{
\begin{tabular}{l c c c c c}
\toprule
&\\[-3ex]               
Methods               & EachMovie      & ML-1M         & ML-20M         & Netflix \\[0.2ex]
\hline
&\\[-2ex]
Matrix Factorization                 & 1.402          & 0.990         & 0.979          & 1.073  \\
Transformer   & \textbf{1.184} & \textbf{0.826}& \textbf{0.949} & \textbf{0.852}\\
\bottomrule
\end{tabular}}
\label{table:simulator_RMSE}
\end{table}

%\debug{The simulated online environment is also trained on users' logs, but not on the same data for training the DEERS framework. We test the simulator on users' logs(not the data for training the DEERS framework and simulator), and the experimental results demonstrate that the simulated online environment ahs overall 90\% precision for immediate feedback prediction task. This result suggests that the simulator can accurately simulate the real online environment and predict the online rewrds, which enables us to test our model on it.}

%\debug{We do not recommend repeated items and remove the ones that have already been recommended from the candidate set}

\noindent
\textbf{Metrics}
%When comparing among models, we are interested in two primary quantities: (1) the average reward of each test user and (2) the accuracy of the generated recommendations, which are the major concerns of an interactive recommender system. Thus, 
We use three standard evaluation metrics to measure the recommendation performance during $T$ time-steps of interactions: average reward (RW@T) (i.e., cumulative rewards divided by $T$), precision (PR@T), and recall (RC@T). 
Following~\cite{zou2020nicf, zhou2020knowledge_graph}, the precision and recall were computed by setting the positive items as the ground truth labels. 
We define an item to be positive if the rating provided from the simulator exceeded $5$-points in EachMovie and $4$-points in the other datasets.

\subsection{Experimental Results} \label{section:main_result}

Table~\ref{table:baseline_performance_result} summarizes the performances of all baseline models for each benchmark dataset. Here, we observe the followings:

Out of all RL-based models, DQNR and NICF consistently outperformed traditional recommendation models (i.e., Random, POP, SASRec), which do not consider the long-term utilities of the users.
By observing the poor performance of SQN and SAC, we found that jointly optimizing the next item prediction task is not beneficial.
In addition, DDPGR also struggled to learn an effective recommendation policy due to the inaccessibility to online interactions.
%Since DDPGR is originally proposed to train the model through online interactions, DDPGR might not be an appropriate choice when solely trained from the offline logged dataset.
Surprisingly, our simple baseline model, GreedyRM, achieved the best performance for all metrics for the EachMovie, ML-20M, and Netflix datasets. Though DQNR showed best performance in the ML-1M dataset, GreedyRM achieved competitive results compared with DQNR. GreedyRM aims to recommend items that maximize the immediate one-step reward and is analogous to DQNR when its discount factor $\gamma$ is 0. Therefore, this naturally raises the question of whether putting more weighting on future rewards harms the recommendation quality.

\subsection{Influence of Future Rewards} \label{section:discount_factor}
To further examine the influence of accounting for future rewards, we conducted an ablation study referring to the discount factor $\gamma$. If the long-term effects of user feedback are significant, assigning a higher weight (i.e., large discount factor) will be beneficial. However, if the long-term effect is insignificant, assigning greater weight on future rewards may serve as extra noise rather than as a beneficial learning signal. For all experiments, we varied the discount factor from 0 to 0.99 and fixed the remaining hyper-parameters to each model’s optimal configurations.

\pgfplotsset{compat=1.16}
\begin{figure}[h]
\begin{tikzpicture}
\begin{groupplot}[
      group style={group size=2 by 2, horizontal sep=1.1cm, vertical sep=1.2cm},
      width=4.7cm, height=3.8cm]
    %%%%%%%%%%%%
    %% REW@40 %%
    %%%%%%%%%%%%
    %------------------------------------------------------------%
    \nextgroupplot[
        title={\footnotesize EachMovie},
        title style={yshift=-1.4ex},
        every x tick label/.append style={font=\scriptsize\color{gray!80!black}},
        xmin=-0.05, xmax=1.05,
        xtick={0, 0.2, 0.4, 0.6, 0.8, 1.0},
        xlabel style={yshift=1mm},
        %xticklabels={32, 64, 128, 256, 512, 1024, 2048},
        %log ticks with fixed point,
        xlabel={\footnotesize$\gamma$},   
        every y tick label/.append style={font=\scriptsize\color{gray!80!black}},
        ymin=3.95, ymax=5.2,
        ylabel={\footnotesize RW@40},
        ylabel style={yshift=-1mm},
        tick align=outside,
        tick pos=left,
        legend style={at={($(0,0)+(1cm,1cm)$)},legend columns=4,fill=none,draw=black,anchor=center,align=center, font=\small},
        legend to name = models]
    \addplot[color=gray, thick,
             mark=none, 
             dashed] 
    coordinates {(-0.02, 5.116) (1.02, 5.116)};
    \addplot[color=customblue,
             mark=*,
             mark size=1] 
    coordinates{(0, 5.114)   (0.1, 5.110) (0.2, 5.112) (0.3, 5.095) (0.4, 5.104) 
                (0.5, 5.103) (0.6, 5.101) (0.7, 5.089) (0.8, 5.088) (0.9, 5.079)
                (0.95, 5.077) (0.99, 5.068)};
    \addplot[color=customred,
             mark=*,
             mark size=1] 
    coordinates{(0, 5.115)   (0.1, 5.106) (0.2, 5.102) (0.3, 5.112) (0.4, 5.084) 
                (0.5, 5.098) (0.6, 5.088) (0.7, 5.095) (0.8, 5.090) (0.9, 5.084)
                (0.95, 5.075) (0.99, 5.071)};
    \addplot[color=customyellow,
             mark=*,
             mark size=1] 
    coordinates{(0, 4.501)   (0.1, 4.561) (0.2, 4.379) (0.3, 4.447) (0.4, 4.410) 
                (0.5, 4.389) (0.6, 4.261) (0.7, 4.397) (0.8, 4.438) (0.9, 4.287)
                (0.95, 4.521) (0.99, 4.215)};
    \addplot[color=customgreen,
             mark=*,
             mark size=1] 
    coordinates{(0, 4.263)   (0.1, 4.358) (0.2, 4.233) (0.3, 4.130) (0.4, 4.058) 
                (0.5, 4.048) (0.6, 4.125) (0.7, 4.052) (0.8, 4.074) (0.9, 4.159)
                (0.95, 4.125) (0.99, 4.056)};
    \coordinate (c2) at (rel axis cs:1,1);    
    %------------------------------------------------------------%
    \nextgroupplot[
        title={\footnotesize ML-1M},
        title style={yshift=-1.4ex},
        every x tick label/.append style={font=\scriptsize\color{gray!80!black}},
        xmin=-0.05, xmax=1.05,
        xtick={0, 0.2, 0.4, 0.6, 0.8, 1.0},
        xlabel style={yshift=1mm},
        %xticklabels={32, 64, 128, 256, 512, 1024, 2048},
        %log ticks with fixed point,
        xlabel={\footnotesize$\gamma$},   
        every y tick label/.append style={font=\scriptsize\color{gray!80!black}},
        ymin=3.80, ymax=4.75,
        ylabel={\footnotesize RW@40},
        ylabel style={yshift=-1mm},
        tick align=outside,
        tick pos=left,
        legend style={at={($(0,0)+(1cm,1cm)$)},legend columns=4,fill=none,draw=black,anchor=center,align=center, font=\small},
        legend to name = models]
    \addplot[color=gray, thick,
             mark=none, 
             dashed] 
    coordinates {(-0.02, 4.680) (1.02, 4.680)};
    \addplot[color=customblue,
             mark=*,
             mark size=1] 
    coordinates{(0, 4.673)   (0.1, 4.656) (0.2, 4.670) (0.3, 4.680) (0.4, 4.670) 
                (0.5, 4.679) (0.6, 4.681) (0.7, 4.695) (0.8, 4.689) (0.9, 4.686)
                (0.95, 4.685) (0.99, 4.680)};
    \addplot[color=customred,
             mark=*,
             mark size=1] 
    coordinates{(0, 4.689)   (0.1, 4.676) (0.2, 4.676) (0.3, 4.654) (0.4, 4.670) 
                (0.5, 4.690) (0.6, 4.670) (0.7, 4.679) (0.8, 4.695) (0.9, 4.685)
                (0.95, 4.677) (0.99, 4.675)};
    \addplot[color=customyellow,
             mark=*,
             mark size=1] 
    coordinates{(0, 4.094)   (0.1, 4.138) (0.2, 4.224) (0.3, 4.154) (0.4, 3.981) 
                (0.5, 4.049) (0.6, 4.176) (0.7, 4.136) (0.8, 4.138) (0.9, 4.187)
                (0.95, 4.134) (0.99, 4.147)};
    \addplot[color=customgreen,
             mark=*,
             mark size=1] 
    coordinates{(0, 4.233)   (0.1, 4.130) (0.2, 4.233) (0.3, 4.048) (0.4, 4.225) 
                (0.5, 4.269) (0.6, 4.105) (0.7, 4.082) (0.8, 4.110) (0.9, 4.169)
                (0.95, 3.961) (0.99, 3.885)};
    \coordinate (c2) at (rel axis cs:1,1);
    %------------------------------------------------------------%
    \nextgroupplot[
        title={\footnotesize ML-20M},
        title style={yshift=-1.4ex},
        every x tick label/.append style={font=\scriptsize\color{gray!80!black}},
        xmin=-0.05, xmax=1.05,
        xtick={0, 0.2, 0.4, 0.6, 0.8, 1.0},
        xlabel style={yshift=1mm},
        %xticklabels={32, 64, 128, 256, 512, 1024, 2048},
        %log ticks with fixed point,
        xlabel={\footnotesize$\gamma$},   
        every y tick label/.append style={font=\scriptsize\color{gray!80!black}},
        ymin=3.75, ymax=4.6,
        ylabel={\footnotesize RW@40},
        ylabel style={yshift=-1mm},
        tick align=outside,
        tick pos=left,
        legend style={at={($(0,0)+(1cm,1cm)$)},legend columns=4,fill=none,draw=black,anchor=center,align=center, font=\small},
        legend to name = models]
    \addplot[color=gray, thick,
             mark=none, 
             dashed] 
    coordinates {(-0.02, 4.536) (1.02, 4.536)};
    \addplot[color=customblue,
             mark=*,
             mark size=1] 
    coordinates{(0, 4.531)   (0.1, 4.526) (0.2, 4.531) (0.3, 4.531) (0.4, 4.515) 
                (0.5, 4.533) (0.6, 4.503) (0.7, 4.492) (0.8, 4.481) (0.9, 4.482)
                (0.95, 4.468) (0.99, 4.452)};
    \addplot[color=customred,
             mark=*,
             mark size=1] 
    coordinates{(0, 4.534)   (0.1, 4.530) (0.2, 4.524) (0.3, 4.527) (0.4, 4.522) 
                (0.5, 4.525) (0.6, 4.527) (0.7, 4.505) (0.8, 4.483) (0.9, 4.487)
                (0.95, 4.493) (0.99, 4.466)};
    \addplot[color=customyellow,
             mark=*,
             mark size=1] 
    coordinates{(0, 4.018)   (0.1, 4.149) (0.2, 4.036) (0.3, 4.029) (0.4, 4.117) 
                (0.5, 4.004) (0.6, 4.045) (0.7, 3.964) (0.8, 3.879) (0.9, 4.073)
                (0.95, 3.973) (0.99, 3.852)};
    \addplot[color=customgreen,
             mark=*,
             mark size=1] 
    coordinates{(0, 4.139)   (0.1, 4.189) (0.2, 4.233) (0.3, 4.129) (0.4, 4.017) 
                (0.5, 4.048) (0.6, 4.215) (0.7, 3.864) (0.8, 3.873) (0.9, 4.123)
                (0.95, 3.845) (0.99, 3.820)};
    \coordinate (c1) at (rel axis cs:0,0);
     %------------------------------------------------------------%
    \nextgroupplot[
        title={\footnotesize Netflix},
        title style={yshift=-1.4ex},
        every x tick label/.append style={font=\scriptsize\color{gray!80!black}},
        xmin=-0.05, xmax=1.05,
        xtick={0, 0.2, 0.4, 0.6, 0.8, 1.0},
        xlabel style={yshift=1mm},
        %xticklabels={32, 64, 128, 256, 512, 1024, 2048},
        %log ticks with fixed point,
        xlabel={\footnotesize$\gamma$},   
        every y tick label/.append style={font=\scriptsize\color{gray!80!black}},
        ymin=3.3, ymax=4.75,
        ylabel={\footnotesize RW@40},
        ylabel style={yshift=-1mm},
        tick align=outside,
        tick pos=left,
        legend style={at={($(0,0)+(1cm,1cm)$)},legend columns=5,fill=none,draw=black,anchor=center,align=center, font=\small},
        legend to name = models]
    \addplot[color=gray, thick,
             mark=none, 
             dashed] 
    coordinates {(-0.02, 4.689) (1.02, 4.689)};
    \addplot[color=customblue,
             mark=*,
             mark size=1] 
    coordinates{(0, 4.655)   (0.1, 4.644) (0.2, 4.626) (0.3, 4.646) (0.4, 4.631) 
                (0.5, 4.615) (0.6, 4.633) (0.7, 4.603) (0.8, 4.592) (0.9, 4.581)
                (0.95, 4.608) (0.99, 4.583)};
    \addplot[color=customred,
             mark=*,
             mark size=1] 
    coordinates{(0, 4.662)   (0.1, 4.663) (0.2, 4.667) (0.3, 4.649) (0.4, 4.658) 
                (0.5, 4.659) (0.6, 4.652) (0.7, 4.654) (0.8, 4.645) (0.9, 4.655)
                (0.95, 4.655) (0.99, 4.519)};
    \addplot[color=customyellow,
             mark=*,
             mark size=1] 
    coordinates{(0, 3.659)   (0.1, 3.629) (0.2, 3.489) (0.3, 3.422) (0.4, 3.497) 
                (0.5, 3.589) (0.6, 3.602) (0.7, 3.456) (0.8, 3.481) (0.9, 3.522)
                (0.95, 3.497) (0.99, 3.487)};
    \addplot[color=customgreen,
             mark=*,
             mark size=1] 
    coordinates{(0, 3.606)   (0.1, 3.546) (0.2, 3.555) (0.3, 3.616) (0.4, 3.703) 
                (0.5, 3.751) (0.6, 3.542) (0.7, 3.503) (0.8, 3.516) (0.9, 3.496)
                (0.95, 3.589) (0.99, 3.605)};
    \addlegendentry{GreedyRM}; 
    \addlegendentry{DQNR}; 
    \addlegendentry{NICF}; 
    \addlegendentry{SQN}; 
    \addlegendentry{SAC}; 
    \coordinate (c2) at (rel axis cs:1,1);
    %------------------------------------------------------------%
    %------------------------------------------------------------%
    %------------------------------------------------------------%
    \end{groupplot}
    \coordinate (c3) at ($(c1)!.5!(c2)$);
    \node[above] at (c3 |- current bounding box.north)
    {\pgfplotslegendfromname{models}};
\end{tikzpicture}
\captionsetup{justification=centering}
\captionof{figure}{Performance comparison of RL-based recommendation models with varying discount factor $\gamma$.}
\label{fig:gamma}
\end{figure}
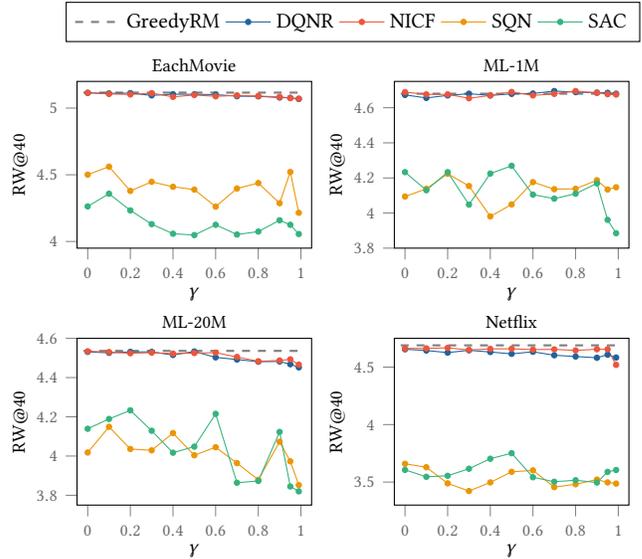

Figure~\ref{fig:gamma} displays the average reward (i.e., RW@40) with respect to each discount factor. We observe that RL-based models display a gradual decrease in cumulative rewards as the models put more weight on future rewards.
Thus, we suspect that long-term effects merely exist between the user feedbacks in the benchmark review datasets.
If this were true, a simple greedy one-step recommendation model will be able to directly maximize the cumulative rewards.
Therefore, we further investigate whether a greedy algorithm can maximize the cumulative rewards in the review datasets.
%by analyzing their long-term effects.

%Here, we hypothesize that GreedyRM is sufficient to learn the recommendation policy that maximizes the cumulative rewards. Thus, accounting more weights on future rewards may serve as an extra noise rather than a beneficial learning signal.
%therefore resulting in the gradual performance decrease of the RL-based models. 
%To verify our assertion, we further investigate whether such greedy algorithm is sufficient to maximize the cumulative rewards under the current experimental settings. 

%all baseline models show a gradual performance decrease for all benchmark datasets. This suggests that focusing on the long-term rewards does not benefit the learning of the agent's policy in optimizing the cumulative reward since the extra information injecting when learning the values of each state acts as an extra noise rather than a beneficial learning signal. 
%We speculate that this may indicate that a greedy policy (i.e., $\gamma = 0$) might be acting already as an optimal policy for the benchmark review datasets.

\subsection{Analyzing Long-Term Effects in Datasets}
Here, we investigated the significance of the long-term effects in the public review dataset.
By comparing the performance between the greedy- and optimal- recommendation policies, we were able to verify whether the greedy algorithm could maximize the cumulative rewards (i.e., no long-term effects in the dataset). 
The performance of the optimal policy can be measured by searching across all actions with the simulator until the last time-step and selecting the best action sequence that maximizes the cumulative rewards.
%If we can observe that the greedy recommendation policy and optimal recommendation policy have equivalent performance, this indicates that the dataset has mere long-term effects. However, computing the optimal policy by training is intractable. 
%greedy하게 추천하는 policy와 optimal policy의 성능이 같으면 long term effect가 없음.
%그런데, optimal policy는 training단계에서는 untractable하다. 
%만약 greedy한 추천이 cumulative reward를 maximize 할 수 있다는 것을 보인다면, dataset에 long term effect가 존재하는지 확인할 수 있다. 
%By comparing the recommendation performance of the optimal policy and greedy policy obtained by search algorithms on the environment simulator, .
%If there exist no gap in the performance, GreedyRM is able to learn the optimal policy and is sufficient to maximize the cumulative rewards.
%\debug{To examine whether the greedy recommendation is sufficient to maximize the cumulative rewards, we compare the ground truth performance between the optimal- and greedy- policy. 
%The performance of the optimal policy can be measured by searching over all actions from the simulator until the last timestep. one can compare the recommendation performance obtained from the optimal- and greedy- policy within the environment simulator.}
%of the ground truth optimal- and greedy-search policy of the environment simulator.}
%between the simulator based optimal- and greedy- search policy.}
%The performance of the optimal policy can be measured by searching over all actions from the simulator until the last timestep.
%optimal- and greedy- search policy of the simulator.
However, this measurement is computationally infeasible since it requires a search across all possible actions for 40 time steps (i.e., $|A|^{40}$).

Therefore, to approximate the optimal recommendation performance using the simulator, we adopt beam search~\cite{beam1977raj}, a commonly used decoding algorithm to generate the sequence that considers the long-term effects ~\cite{freitag2017beam}.
A beam search maintains a beam of $k$ possible trajectories, updating them incrementally by ranking their extensions via the objective score.
Since the memory requirements for beam search is proportional to the action size and the review dataset has a large number of actions (e.g., 17,769 for ML-20M), we use beam size $k \in \{1,10\}$. 
Here, we note that the beam search ($k=1$) is identical to the greedy policy.
%obtained from beam search ($k=1$) is identical to approximate the optimal policy. 

%Is is a common protocol in the machine translation task ~\cite{GPT} to apply beam search rather than performing a greedy search. In this experiment, we search whether there is a performance gap between a greedy search and a beam search approach. The beam search approach represents an optimal strategy when generating long sequences. 

%We find that there is only a small gap in performance between the greedy search approach and bea search approach. This verifies that acting greedily is able to produce near-optimal results for the review dataset. This indicates that a greedy approach such as GreedyRM is sufficient enough to model the user-item interactions and produce high valued recommendations. 

\begin{table}[h]
\caption{Relative performance of the greedy policy against the performance of the beam search at $k=10$.}
\begin{tabular}{l c c c c c}
\toprule
&\\[-3ex]               
Beam size           & EachMovie  & ML-1M    & ML-20M     & Netflix       \\[0.2ex]
\hline
&\\[-2ex]
$k=1$     & 0.9960     & 0.9844   & 0.9981     & 0.9974        \\
\bottomrule
\end{tabular}
\label{table:greedy_beam}
\end{table}

Table \ref{table:greedy_beam} reports the relative recommendation performance of the greedy search (i.e., beam search at $k=1$) with respect to the beam search performance at $k=10$. We found that the greedy search matches the performance of the beam search by more than 99.5\% for EachMovie, ML-20M, and Netflix dataset. 
This implies that the benefits of considering the long-term effects are marginal and provides an explanation for the high performance of GreedyRM for these datasets in Table~\ref{table:baseline_performance_result}.
We also observed a slight performance gap between the greedy search and the beam search in ML-1M. This indicates that considering the long-term effects can be slightly beneficial for ML-1M, which matches with our experimental findings in Table~\ref{table:baseline_performance_result} where DQNR achieved the best performance in ML-1M.

%pointing out some notions of long-term rewards for ML-1M. 
%Therefore we believe that the benefits of the recommendation that considers the long-term effects are marginal in the review dataset.
%In addition, greedy approach is sufficient to maximize the cumulative rewards under the current experimental settings.

\section{Conclusion and Discussion} \label{section:conclusion}

Recently, a lot of community effort in IRS has been devoted to develop the RL-based recommendation algorithms to effectively model the long-term effects between the recommendations.
However, our findings imply that the benefits of accounting for the long-term effects can be marginal in the public review dataset.
Therefore, to accurately benchmark the performance of the RL-based recommendation algorithms, it is crucial to validate the significance of the long-term effects prior to the evaluation.
We suggest that an evaluation protocol should (i) perform a dataset validation procedure by comparing beam- and greedy-search performance to verify the existence of beneficial long-term rewards; and (ii) include a simple reward model that greedily selects items as a baseline. We will make our code publicly available to ensure the reproducibility of our work. 

\section*{Acknowledgments}
This work was supported by Institute of Information \& communications Technology Planning $\&$ Evaluation (IITP) grant funded by the Korea government(MSIT) (No. 2019-0-00075, Artificial Intelligence Graduate School Program(KAIST), and No. 2020-0-00368, A NeuralSymbolic Model for Knowledge Acquisition and
Inference Techniques).

%\newpage
%\appendix
%\input{tex/Appendix}

%\begin{acks}
%To Robert, for the bagels and explaining CMYK and color spaces.
%\end{acks}

\balance
\bibliographystyle{ACM-Reference-Format}
\bibliography{reference}
\clearpage
% \appendix
% \input{tex/09 Appendix}

\end{document}